\documentclass[prb,aps,twocolumn]{revtex4}
\usepackage{graphicx}
\newcommand{\normwidth}{0.5\columnwidth}

\newcommand{\bphi}{{{\mbox{\boldmath $\phi$}}}}

\begin{document}

\title{Thermal transport in one-dimensional spin gap systems}
\author{E. Orignac}
\affiliation{Laboratoire de Physique Th{\'e}orique de l'{\'E}cole Normale
  Sup{\'e}rieure CNRS-UMR8549 24,Rue Lhomond 75231 Paris Cedex 05
  France}
\author{R. Chitra}
\affiliation{Laboratoire de Physique Th{\'e}orique des Liquides
CNRS-UMR7600 \\
  Universit{\'e} Pierre et Marie Curie, 4  Place
  Jussieu 75252 Paris Cedex 05 France}
\author{R. Citro}
\affiliation{Dipartimento di Fisica ``E. R. Caianiello'' and
Unit{\`a} I.N.F.M. di Salerno\\ Universit{\`a} di Salerno, Via S.
Allende, 84081 Baronissi (Sa), Italy}

\begin{abstract}
We study thermal transport in one dimensional spin systems both in the
presence and absence of impurities.
In the absence of disorder, all these spin systems display
a temperature dependent Drude peak in  the thermal conductivity.
In gapless systems, the low temperature Drude weight is
proportional to temperature and
to the central charge which characterizes the conformal field theory
that describes the system at low energies.
On the other hand, the low temperature Drude weight of spin gap
systems shows an
activated behavior modulated by a power law.
For temperatures higher than the spin gap, one recovers the linear $T$
behavior akin to gapless systems. For temperatures larger than the
exchange coupling, the Drude weight decays as $T^{-2}$. We argue
that this behavior is a generic feature of quasi one dimensional
spin gap systems with a relativistic-like 
low energy dispersion. We  also consider  the effect of a magnetic field
on the Drude weight with emphasis on the
commensurate-incommensurate transition induced by it.
We then study the effect of nonmagnetic impurities on the
thermal conductivity of the dimerized XY chain and the spin-$\frac12$
two leg ladder. Impurities destroy the Drude peak and the thermal
conductivity  exhibits a purely activated behavior at low
temperature, with an activation gap renormalized by disorder.
The relevance of these results for experiments is briefly discussed.

\end{abstract}
\maketitle

\section{Introduction}
The past many years have seen a resurgence of interest, both
 theoretical and experimental in
 quasi-one dimensional spin  gap systems.
  Well known examples of systems with a gap are  spin chains with dimerization,
frustration
and anisotropy\cite{haldane_dimerized,nijs_equivalence}. 
Another interesting example is the
  two leg spin $S=\frac12$ ladder which  was proposed as a toy model  for
the
pseudogap phase
 in  high temperature
 superconductors.\cite{rice_srcuo} Renewed  interest in these systems
 was triggered by the availability of anisotropic
 materials\cite{takano_spingap,chaboussant_cuhpcl,uchara_SrCaCuO} in
 which the magnetic properties of the insulating phase could be
 ascribed to one or quasi-one dimensional spin systems. 
The dynamical properties  of these
quasi-1d spin phases have been extensively studied using
standard techniques like   neutron scattering and NMR
 measurements.\cite{dagotto_2ch_review,dagotto_supra_ladder_review}
More recently, heat transport is being used as a complementary probe
to study low dimensional spin
 systems.  Measurements of thermal conductivity have been carried out
in systems such as the spin chain materials
 $\mathrm{SrCuO_2}$ and
 $\mathrm{Sr_2CuO_3}$\cite{sologubenko_therm_sr2cuo3,sologubenko_spinchains_th},
 the spin Peierls system
 $\mathrm{CuGeO_3}$\cite{ando_sP_thermal}
and  the spin ladder materials $\mathrm{(Sr,Ca,La)_{14}Cu_{24}O_{41}}$
 \cite{sologubenko_thermal_ladder,hess_thermal_ladder,kudo_thermal_ladder}.  The huge
anisotropy seen in the thermal conductivity in the
directions parallel and perpendicular to the chains or the ladders, indicates
that magnetic excitations of these quasi-1D systems do play an
 important role in heat transport. This is further confirmed by
 measurements in the presence of a magnetic
field\cite{ando_sP_thermal}.

Various attempts have been made to extract from these measurements, the purely magnetic
contribution to
 the thermal conductivity. This is often done by subtracting a phonon
background  calculated within a Debye model.
 However, in order to account for
 the entire  magnetic contribution,
one needs to understand the interactions of the spin
 excitations of the low dimensional spin systems with themselves and, with defects and
 phonons. This is a non-trivial problem since the spin excitations
 are not  necessarily weakly interacting, and the form of interaction of these spin
 excitations with phonons or defects is usually rather complicated.
Consequently, experimental results have been fitted using various
 phenomenological kinetic theory expressions for non-interacting
 spinons or magnons.
This effort to obtain the purely magnetic contribution to the thermal
conductivity has   stimulated theoretical studies of
 thermal transport in spin chain and spin ladder
 systems.\cite{alvarez_ladder,saito_thermal,heidrich_frustrated}
In the absence of extrinsic scattering such as phonons or defects,
 some studies\cite{alvarez_ladder} showed that the frequency dependent thermal
 conductivity $\kappa(\omega,T)=\pi\tilde{\kappa}(T)\delta(\omega)$,
 where $\tilde{\kappa}(T)$ is the thermal Drude weight. 
 However, the  Drude weight 
extracted from finite size zig-zag ladders \cite{heidrich_frustrated}
 seems to be at odds with
the idea of an infinite thermal conductivity in spin systems without
disorder.

\begin{figure}
\centerline{\includegraphics[angle=-0,width=\normwidth]{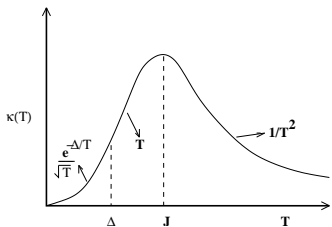}}
\caption{Schematic representation of the thermal Drude
weight$\tilde{\kappa}$  in spin gap systems without impurities}
\label{fig:weight}
\end{figure}

In this paper, we use analytical methods to  revisit the problem of the thermal
conductivity for various quasi 1d spin systems with special emphasis
on the two leg spin-$\frac12$ ladder. In the absence of impurities, we
present results which should be valid for spin gap systems possessing 
low energy triplet excitations and gapless
systems irrespective of the details of the nature of the interaction.
A schematic representation of the thermal Drude weight for a spin gap
system is shown in Fig.\ref{fig:weight}.
We also study the effect of one/many impurities on the thermal
conductivity of the  spin ladder and show that
 in the presence of impurities, the thermal conductivity
is not simply given by the Drude weight  times a temperature
independent scattering time.

The paper is organized as follows:
in Sec.~\ref{sec:pure-case}, we discuss the temperature dependence
 of the Drude weight for different spin systems ranging from  gapless
integrable spin chains
to spin gap systems like the spin ladder  and the spin$-1$ chain.
In Sec.~\ref{sec:disorder-case}, we discuss the effect of
 impurities on the thermal conduction. In particular, we use the
Landauer approach \cite{landauer_formula} to  evaluate
 the effect of a single non-magnetic  impurity on the thermal
conductivity of the  ladder and the XY-chain.
 We then study the
 effect of a finite concentration of impurities on the ladder.
Finally,  we present a comparison of  our results to experiments and other theoretical
 work on the subject.

\section{Translationally invariant  systems}\label{sec:pure-case}

In this section, we  briefly outline the general definitions of the thermal
current and the thermal conductivity calculated within linear
response theory. We then use this formalism to calculate the dc thermal
conductivity
of various spin systems  with and without a gap to low energy
excitations. The examples considered are: gapless integrable spin
chains described by a conformal fixed point, the spin-$\frac12$ ladder,
the dimerized XY chain  and lastly
the case of massive bosons. We also consider the
effect
of a magnetic field on the thermal Drude weight.

\subsection{Definition of thermal current and thermal conductivity}

We consider a system defined by a Hamiltonian density ${\cal H}(x)$ so
that the total Hamiltonian is $H=\int dx {\cal H}(x)$.
Conservation of energy leads to the continuity equation
\begin{eqnarray}
  \label{eq:continuity}
  \partial_t {\cal H}(x,t)+\partial_x j_e(x,t)=0,
\end{eqnarray}
\noindent where $j_e$ is the energy(thermal) current of the system. In
the absence of charged excitations, the energy and thermal current are
equivalent.
 Eq.(\ref{eq:continuity})  permits a definition of the thermal
current in terms of  the Hamiltonian density.
Within linear response theory\cite{luttinger_thermal},
the energy current response function at
temperature $T>0$ reads:
\begin{equation}\label{eq:response-def}
\chi(\omega,T)= \int dx \int_0^\infty dt e^{i\omega t} \langle
[j_e(x,t), j_e(0,0)] \rangle,
\end{equation}
\noindent where $\langle \ldots \rangle$ indicates both quantum and
thermal averaging.
 It is often easier to  use the imaginary time formalism
to calculate   $\chi(i\omega_n)$ where $\omega_n$ are the Matsubara frequencies
and then analytically continue to real frequencies $i\omega_n\to \omega +i0$ to
obtain  $\chi(\omega,T)$.\cite{mahan_book}
  The frequency dependent thermal conductivity is
then given by \cite{luttinger_thermal}:
\begin{equation}\label{eq:conduc-def}
\kappa(\omega,T)=\frac 1 {i \omega T} \left[ \chi(0,T) -
\chi(\omega,T)\right],
\end{equation}
\noindent
In general, in the absence of phonons or impurities, the total thermal current
$J_e(t)=\int dx j_e(x,t)$ is conserved.  This conservation permits an
alternative but equivalent  formulation of the thermal conductivity
(\ref{eq:conduc-def})
\begin{equation}
\kappa(\omega,T)= \frac 1 {2 L T^2} \int_0^\infty \langle \{
J_e(t),J_e(0)\} \rangle e^{i\omega t} dt,
\end{equation}
\noindent where $L$ is the system size. Since $J_e$ is conserved, the
total current is time independent and
\begin{equation}\label{eq:drude-thermal}
\kappa(\omega,T)= \frac{\pi}{L T^2} \langle J_e^2 \rangle \delta(\omega) =
\tilde{\kappa}(T) \delta(\omega).
\end{equation}
This implies an infinite dc thermal conductivity,
with a temperature dependent  Drude weight $\tilde{\kappa}(T)$ which
vanishes at zero temperature. This
thermal Drude weight  has
been studied numerically for some spin gap systems in
Refs.\onlinecite{alvarez_ladder,heidrich_frustrated}. In the following
sections, we present an analytical  discussion of the behavior of the Drude
weight in various gapless  and  gapped quasi-one dimensional spin systems.

\subsection{ Gapless Integrable Spin
  Chains}\label{sec:pure-conformal}
In this section, we present results for the thermal Drude weight of
integrable spin chains which are characterized by a vanishing
singlet-triplet
gap.
One example of such a system, is
 the integrable spin-$\frac12$ Heisenberg model, which is known to
have
 a Drude
weight
that vanishes linearly as temperature goes to zero ${\tilde \kappa}= \pi^2vT/3$
\cite{kluemper_xxz}, where $v$ is the velocity of spin excitations. Other interesting systems,  are the
various integrable generalizations of the spin-$\frac 12$ Heisenberg
spin chain, like the
  spin$-S$ chain
models\cite{takhtajan_spin_s,babujian_spin_s} and  $SU(N)$
spin chain models\cite{sutherland_su3_ba,uimin_su3_ba,lai_su3_ba}.
 This
description allows one to
easily  obtain    the low temperature  thermal conductivity of these
chains.
The long wavelength behavior of these  integrable
 systems are described by a
conformal invariant fixed point
 \cite{witten_wz,affleck_wz,affleck_strongcoupl}.  This
description allows one to
easily  obtain    the low temperature  thermal  Drude weight of these
chains.
 The effective Hamiltonian of these systems
has the generic form
\begin{eqnarray}
  \label{eq:wz-models}
  H=\int dx [{\cal H}_R(x)+{\cal H}_L(x)]
\end{eqnarray}
where ${\cal H}_R$ and ${\cal H}_L$ describe right and left moving chiral modes.
In addition,  chirality imposes the constraints ${\cal H}_R(x,t)={\cal H}_R(x-vt)$ and
${\cal H}_L(x,t)={\cal H}_L(x+vt)$. This leads to the
following relation
\begin{eqnarray}
  \label{eq:contin-wz}
  \partial_t ({\cal H}_R(x,t)+{\cal H}_L(x,t))=-v\partial_x({\cal
    H}_R(x,t)-{\cal H}_L(x,t))
\end{eqnarray}
which results in a  thermal current density:
\begin{eqnarray}
  \label{eq:energy-curr-wz}
  J_e=v \int dx [:{\cal H}_R(x):-:{\cal H}_L(x):]
\end{eqnarray}
\noindent
As before, since  $[H,J_e]=0$, the Drude weight is given by
\begin{eqnarray}
  \label{eq:drude-wz}
  \tilde{\kappa}(T)=\frac{\pi}{L T^2} \langle J_e^2 \rangle
\end{eqnarray}
Moreover, since there is no interaction between the right and the left moving
modes,  $\langle J_e^2 \rangle =v^2 \langle H^2 \rangle$.
One thus immediately obtains the result $\tilde{\kappa}_{WZ}(T)=\pi C_v(T)
v^2$, where $C_v$ is the specific heat of these modes.
For conformally invariant modes with a central charge $c$, the
specific heat  is given
by $C_v(T)=\frac{\pi T}{3v}c$, leading
to a thermal Drude weight:
\begin{eqnarray}
  \label{eq:weight-conformal}
  \tilde{\kappa}(T)=\frac{\pi^2 Tv}{3}c
\end{eqnarray}
For the integrable spin-1 chain at the Takhtajan-Babujian point \cite{takhtajan_spin_s,babujian_spin_s}, this
weight can also be recovered from  explicit calculations using the
Majorana formalism  to be discussed in the forthcoming sections.
 For a theory
described by a free massless boson like the spin-$\frac12$ Heisenberg
chain,
which has a central charge $c=1$,
  this weight is $\pi^2 Tv/3$, which can also be checked by direct
calculations of the thermal susceptibility.\cite{heidrich_frustrated}
For systems with a Luttinger liquid like description\cite{schulz_houches_revue} with  a Luttinger
exponent $K$,  the present derivation illustrates clearly that the weight
$\tilde{\kappa}$ is independent of the  Luttinger exponent or equivalently,
the compactification radius of the free bosonic Luttinger
field. Considering  the case of $XXZ$ chains, this result implies that
the
thermal Drude weight is independent of the anisotropy $J_z/J_{xy}$
which is in agreement with Bethe Ansatz calculations
on the XXZ spin chain in the Luttinger liquid
regime.\cite{kluemper_xxz}

\subsection{Spin-$\frac 12$ ladder}\label{sec:pure-ladder}
Here  and in the following sections, we focus exclusively on spin
gap systems.
We first apply the  formalism of  Sec.~\ref{sec:pure-case}
to the clean two leg spin ladder. The Hamiltonian of the two leg spin
ladder is
\begin{eqnarray}
  \label{eq:lattice-hamiltonian}
  H=J_\parallel \sum_{i\atop {p=1,2}} {\bf S}_{i,p} \cdot {\bf
  S}_{i+1,p} + J_\perp \sum_i {\bf S}_{i,1}\cdot {\bf S}_{i,2},
\end{eqnarray}
\noindent where the ${\bf S}_{i,p}$ are spin-$\frac 12$ operators, and
the exchange constants $J_\parallel,J_\perp>0$. For weak interchain coupling $J_\perp\ll
J_\parallel$, the spin ladder can be described by a continuum theory of
spinless Majorana fermions \cite{shelton_spin_ladders}. The continuum
Hamiltonian
reads:
\begin{eqnarray}
  \label{eq:continuum-hamiltonian}
H&=&\sum_{a=0}^3 \int dx {\cal H}^a(x),\\ 
{\cal H}^a(x)&=&
\frac{-iv}{4}\lbrack \xi^a_R(x) \partial_x \xi^a_R(x)-(\partial_x
\xi^a_R)(x) \xi^a_R(x)\nonumber \\ & & -\xi^a_L(x)
\partial_x \xi^a_L(x) + (\partial_x \xi^a_L)(x)\xi^a_L(x)\rbrack +i m^a \xi^a_R(x)\xi^a_L(x)
\end{eqnarray}

\noindent where the velocity of the Majorana fermions  $v=\frac \pi 2
J_\parallel a$  ($a$ is the lattice spacing).
Physically,  the Majorana modes $\xi^a_{R,L}$ $(a=1,2,3)$  with masses
$m_{1,2,3}=J_\perp/(2\pi)\equiv \Delta$
 describe triplet excitations  with a gap $\Delta$   and $\xi^0_{R,L}$ with mass  $m_0=-3J_\perp/(2\pi)=-3\Delta$ describe singlet excitations.
 We remark that the
bosonized version of the  low energy Hamiltonian~(\ref{eq:continuum-hamiltonian})
describes more general spin ladder models than the one considered in
(\ref{eq:lattice-hamiltonian}).\cite{nersesyan_biquadratic,kim}
  Using (\ref{eq:continuity}),  the energy
current for the ladder takes the form
\begin{eqnarray}\label{eq:majorana-therm-curr}
j_e(x)&=&\sum_{a=0}^3 j_e^a(x), \\ j_e^a(x)&=& \frac{-iv^2}{4} \left[
\xi^a_R \partial_x \xi^a_R - (\partial_x \xi^a_R ) \xi^a_R + \xi^a_L
\partial_x \xi^a_L- (\partial_x \xi^a_L ) \xi^a_L \right].\nonumber
\end{eqnarray}

From (\ref{eq:majorana-therm-curr}) and (\ref{eq:drude-thermal}),
the total Drude weight for the spin ladder is found to be
\begin{equation}
\tilde{\kappa}(T)=\sum_a
\tilde{\kappa}^a(T)=\tilde{\kappa}^0(T)+3\tilde{\kappa}^1(T).
\end{equation}
Since the Majorana fermions are essentially free, the
correspondence between Majorana and Dirac fermions can be  used to
evaluate  the Drude weight $\tilde{\kappa}^a(T)$
\begin{equation}\label{eq:kappa-t}
\tilde{\kappa}^a(T)= \frac 1 {8T^2} \int_{-\Lambda }^{\Lambda } dk \frac
{v^4 k^2}{\cosh^2 \left( \frac{\epsilon_a(k)}{2T}\right) },
\end{equation}
where the energy dispersion $\epsilon_a(k)=\sqrt{(vk)^2+m_a^2}$ and $\Lambda \sim
\frac {2\pi}{a}$ is the lattice induced  ultra-violet cutoff.
The details of the calculation are presented in
Appendix~\ref{app:weight}.
The thermal Drude weight of the spin ladder is now given by
\begin{eqnarray}
  \label{eq:ladder-kappa}
  \tilde{\kappa}(T)= \frac 3 {8T^2} \int_{-\Lambda }^{\Lambda } dk \frac
{v^4 k^2}{\cosh^2 \left( \frac{\epsilon_1(k)}{2T}\right) } +
\frac 1 {8T^2} \int_{-\Lambda }^{\Lambda } dk \frac {v^4 k^2}{\cosh^2
\left( \frac{\epsilon_0(k)}{2T}\right) }
\end{eqnarray}
\noindent
At low enough  temperatures
$T\ll J_\perp/(2\pi)$,  the triplet excitations   are the dominant carriers of
heat and
\begin{equation}\label{eq:tcond-low}
\tilde{\kappa}(T) = 3 \sqrt{\frac{\pi \Delta^2} {2T}} v
e^{-\Delta/T}
\end{equation}
For temperatures  $\Delta\ll T\ll  J_{\parallel}$, the coupling between the two spin
half chains becomes irrelevant and we recover  $\tilde{\kappa} \propto T $ i.e., it  is the sum of the
Drude
weights of two independent spin$-\frac12$ chains cf. Sec.~\ref{sec:pure-conformal}. For temperatures $T\gg J_{\parallel}$, the temperature dependence in the
integrand of (\ref{eq:ladder-kappa}) becomes negligible and since the
$k$ integral is bounded on a lattice,
the
thermal Drude weight decays as
${\tilde \kappa} \propto T^{-2}$. The prefactor depends on the cutoff $\Lambda$ and
it is reasonable to assume that the continuum theory over-estimates
this
prefactor.
To summarize,  the Drude weight of the spin ladder has three regimes :i) at
very low temperatures, $T\ll \Delta$, we obtain the exponential
behavior (\ref{eq:tcond-low}) ii) for intermediate temperatures, $\Delta \ll T \ll
J_\parallel$, $\tilde{\kappa}\propto T$ and  iii) for $T\gg J_{\parallel}$,$\tilde{\kappa} \sim
1/T^2$. Since  $\tilde{\kappa}(T \to 0)=0$, this implies
the presence of at least one maximum in $\tilde{\kappa}$ at a finite
temperature for a lattice model. We expect $\tilde{\kappa}$ to have a peak in the vicinity of
 $T\sim J_\parallel$ (cf. Fig.\ref{fig:weight}).
We note that  the  numerical results for  $\tilde{\kappa}(T)$ for the
ladder presented in Ref.\onlinecite{alvarez_ladder} confirm our
picture.

To study the effect of an applied magnetic field $h$ on the thermal
conductivity, we first note that the effect of the magnetic field is
to alter the dispersion of the triplet.  The degenerate triplet
dispersion $\epsilon_1(k)$ now splits into three branches $\epsilon_1(k)
+h$,$\epsilon_1(k)$ and $\epsilon_1(k)-h$ and the singlet dispersion $\epsilon_0(k)$ remains
unaltered.
The Drude weight in the presence of the field is now given by:
\begin{widetext}
\begin{equation}\label{eq:weight-field}
\tilde{\kappa} = \frac 1 {8 T^2} \int_{- \Lambda }^{\Lambda } dk
\left(\frac{\partial \epsilon_1(k)}{\partial k}\right)^2
\left\{ \frac {(\epsilon_1(k)-h)^2} {\cosh^2\left(\frac{\epsilon_1(k) - h}{2T}\right)}
+\frac {(\epsilon_1(k) + h)^2} {\cosh^2\left(\frac{\epsilon_1(k) + h}{2T}\right)} + \frac {\epsilon_1(k)^2}
{\cosh^2\left(\frac{\epsilon_1(k)}{2T}\right)} \right\} + \frac 1 {8
T^2} \int_{- \Lambda }^{\Lambda } dk  \frac {v^4 k^2}
{\cosh^2\left(\frac{\epsilon_0(k)}{2T}\right)}
\end{equation}
\end{widetext}
\noindent
There are now two regimes of interest:  $h \ll  \Delta$ and  $h > \Delta$.  In the former case, the effective gap $\Delta-h$
dominates the thermal conductivity and $\tilde{\kappa}\propto
e^{-(\Delta-h)/T}/\sqrt{T}$.  The magnetic field leads to  a sufficient enhancement of
the  low temperature thermal conductivity.  The physical reason is that the increase of
the number of triplet excitations with $S^z=+1$ strongly dominates
the diminution of the number of excitations having $S^z=-1$. For
$h\sim \Delta$, the dispersion of the 
Majorana fermions  describing the $S_z=+1$ sector is no
longer relativistic-like  but quadratic,  $\epsilon(k) \propto k^2$,
resulting in  $\tilde{\kappa} \sim T^{3/2}$. Finally, for
$h>\Delta$, the gap in spin ladder is
closed\cite{chitra_spinchains_field,sachdev_qaf_magfield}, and the
fermionic excitations have an effective linear dispersion,
leading to $\tilde{\kappa}(T)=\frac{\pi^2 T}{3} {\tilde v}(h)$, where
 the effective Fermi velocity ${\tilde v}(h)=v \sqrt{1-(\Delta/h)^2}$.

Let us note that the above results
(\ref{eq:ladder-kappa}),~(\ref{eq:weight-field}) are also relevant for
spin-1 chains. Indeed, spin-1 chains are also described at low energy
by massive Majorana fermions, with a Hamiltonian similar to
(\ref{eq:continuum-hamiltonian}), except that the singlet mode $\xi^0$ is
absent\cite{tsvelik_field}. This mapping to  massive Majorana fermions
originally derived for a spin-1 chain with bi-quadratic interactions in the
vicinity of the Takhtajan-Babujian
point\cite{takhtajan_spin_s,babujian_spin_s},  is also expected to
 provide a  qualitatively description of the low energy properties
of the Heisenberg spin-1 chain.
Therefore, the thermal
conductivity of the spin-1 chain is easily obtained by  taking the limit $\epsilon_0 \to
\infty$ in Eqs.~(\ref{eq:kappa-t}) and (\ref{eq:weight-field}).
For the spin-1 chain,
the low temperature behavior of the Drude weight in the thermal
conductivity is still given by (\ref{eq:tcond-low}).  The main
difference between the ladder and the spin-1 chain stems from the fact
that while in the former the gap to triplet excitations  is small, in the latter the gap
$\Delta$ is of the order of the Heisenberg exchange $J$ ($\Delta=0.41 J$).
 Consequently, the intermediate
regime of linear temperature dependence of $\tilde{\kappa}$
can hardly be observed  in the spin-1 chain. However,
reasonably strong  bi-quadratic interactions can reduce the gap
appreciably  rendering an observation of an intermediate linear regime
possible.  This  predicted linear behavior might in fact be observable
in  the compound
$\mathrm{LiV_2GeO_6}$ which is expected to have sizeable biquadratic
interactions\cite{millet_biquad}.

\subsection{Dimerized XY Chain} \label{sec:pure-XY}
We consider a spin-$\frac 12 $  XY chain with alternating exchange in
an external magnetic field $h$, described by the
Hamiltonian:
\begin{eqnarray}
  \label{eq:XYchain-ham}
  H&=&\sum_n J_1 (S_{2n}^xS_{2n+1}^x+S_{2n}^yS_{2n+1}^y)\nonumber \\
  && + \sum_n J_2 (S_{2n}^xS_{2n-1}^x+S_{2n}^yS_{2n-1}^y) -h S_n^z
\end{eqnarray}
Using the Jordan-Wigner transformation\cite{jordan_transformation},
\begin{eqnarray}
  \label{eq:jordan-wigner}
  S^+_{n}&=&a^\dagger_n \cos\left(\sum_{m<n} a^\dagger_m a_m\right),\nonumber \\
  S^z_n&=&a^\dagger_n a_n-\frac 1 2,
\end{eqnarray}
\noindent where the $a,a^\dagger$ are fermion annihilation and
 creation
operators,  the Hamiltonian (\ref{eq:XYchain-ham})  can be rewritten
 as:
\begin{eqnarray}
  \label{eq:XY-fermionized}
  H=J_1\sum_n (a^\dagger_{2n+1}a_{2n}+ \text{H.c.}) + J_2 \sum_n (a^\dagger_{2n-1}a_{2n}+ \text{H.c.})
\end{eqnarray}
\noindent
Diagonalizing the above Hamiltonian, we obtain
\begin{eqnarray}
  \label{eq:XY-diagonalized}
  H=\sum_k [E(k)-h] a^\dagger_{k,+} a_{k,+} - [E(k)+h]a^\dagger_{k,-}
  a_{k,-},
\end{eqnarray}
\noindent where $E(k)=\sqrt{(J_1-J_2)^2+4 J_1 J_2 \cos^2 k}$. Clearly,
the dimerization induces a gap in the dispersion.
Using the results of  the previous sections and Appendix
\ref{app:mag-clean}, the thermal Drude weight of this dimerized chain is
\begin{eqnarray}
  \label{eq:XY-weight}
  \tilde{\kappa}_{XY}(T,h)&=&\frac 1 {8T^2}  \int_{-\frac \pi a}^{\frac{\pi}{a}}  dk \left[ \frac
    {(E(k)-h)^2}{\cosh^2\left(\frac{E(k)-h}{2T}\right)}\right. \nonumber \\ && \left. +  \frac
    {(E(k)+h)^2}{\cosh^2\left(\frac{E(k)+h}{2T}\right)}\right]
\left(\frac{\partial E(k)}{\partial k}\right)^2
\end{eqnarray}
For $T\ll |J_1-J_2|$, and $|J_1-J_2| \ll \sqrt{J_1 J_2}$
the physics is similar to that  of the continuum model of the
weakly coupled ladder discussed in  Sec.~\ref{sec:pure-ladder}.
The fact that the model is defined on a lattice allows us to  verify
that for  $h=0$ and at very high temperatures the thermal conductivity indeed decays
as
$T^{-2}$.
For $T\gg \sqrt{J_1 J_2}$, and $h=0$, since the energy spectrum
of the XY chain is bounded, one has:
\begin{eqnarray}
  \label{eq:XY-highT}
  \tilde{\kappa}_{XY}(T)=\frac 1 {4T^2}  \int_{-\frac \pi a}^{\frac{\pi}{a}}  dk E(k)^2
  \left(\frac{\partial E(k)}{\partial k}\right)^2
\end{eqnarray}
This result is in fact more general.  Since,  at high temperatures, $\langle
J_e^2 \rangle $
is finite for  a lattice model, we have the asymptotic behavior
${\tilde \kappa} \sim\langle
J_e^2 \rangle_{T=\infty} /T^2 $. Another  interesting limit is when $\sqrt{J_1 J_2}\ll
|J_1-J_2|$, i.e. when the spin gap is much larger than the bandwidth
of magnetic excitations. In this case, for $\sqrt{J_1J_2}\ll
T \ll |J_1-J_2|$, replacing  $E(k)$ in (\ref{eq:XY-weight}) by
$|J_1-J_2|\equiv \Delta$  we obtain
\begin{eqnarray}
  \label{eq:schottky}
  \tilde{\kappa}(T)\simeq \frac{\pi J_1^2 J_2^2} {a T^2 \cosh^2\left(\frac \Delta
      {2T}\right)}
\end{eqnarray}
\noindent
Note that the Drude weight can be recast in the form
 $\tilde{\kappa}(T)=\pi C_v(T)
v_{\text{eff.}}$, where $C_v(T)$ is the specific heat of a fermion
that can occupy two levels separated by $|J_1-J_2|$ and
with an effective velocity $v_{eff.}\sim J_1 J_2/|J_1-J_2|$. However,
this analogy cannot be extended systematically to other spin gapped
systems.

\begin{figure}[htbp]
  \begin{center}
\centerline{\includegraphics[angle=-0,width=\normwidth]{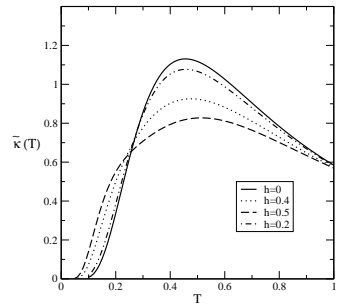}}
    \caption{The field dependence of the thermal Drude weight $\tilde{\kappa}(T)$
       for $|J_1-J_2|=1$, and
      $\sqrt{J_1J_2}=1$.}
    \label{fig:kappa-h}
  \end{center}
\end{figure}

\begin{figure}[htbp]
  \begin{center}
\centerline{\includegraphics[angle=-0,width=\normwidth]{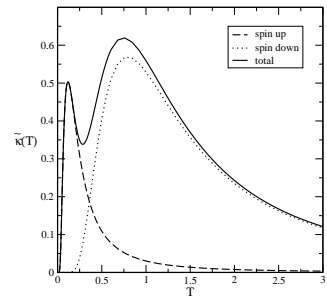}}
    \caption{The double peak structure in $\tilde{\kappa}$ for $h=0.8$,
$|J_1-J_2|=1$ and $\sqrt{J_1 J_2}=1$.}
    \label{fig:double-peak}
  \end{center}
\end{figure}

Turning to the effect of the magnetic field, the Drude weight can
again be rewritten as  (see
 Eq.~(\ref{eq:kinetic-interpretation})) :
\begin{eqnarray}
  \tilde{\kappa}(T,h)= \int_{-\frac \pi a}^{\frac{\pi}{a}} dk [C_v(\epsilon(k)-h)+C_v(\epsilon(k)+h)]v^2(k),
\end{eqnarray}
\noindent where $C_v(\epsilon)$ is the specific heat of a single fermion of
energy $\epsilon$ and the velocity $v(k)= \epsilon \partial\epsilon/ \partial k$. This form helps us
derive a kind of sum rule for the thermal conductivity. For a free fermion, one has:
\begin{eqnarray}
  \int_0^\infty \frac{C_v(T)}{T} dT=S(T=\infty)-S(T=0)=k_B \ln 2,
\end{eqnarray}
And thus:
\begin{eqnarray}
\label{eq:sumrule}
  \int_0^\infty \frac{\tilde{\kappa}(T,h)}{T} dT=k_B \ln 2 \int dk v^2(k).
\end{eqnarray}
We note that the integral is independent of the magnetic field, so we
have a kind of ``sum rule''. Since in the presence of the
magnetic field, it is easily seen that the low temperature thermal weight is enhanced by a
factor $e^{h/T}$, this necessarily implies that
for higher temperatures, the thermal weight must decrease when a
magnetic field is applied This scenario is confirmed by  Fig.~\ref{fig:kappa-h}.
We also note that for high magnetic fields, a double
peak structure appears in the thermal weight as seen in Fig.~\ref{fig:double-peak}.
The double peak results from the low temperature shift of the
maximum of the contribution of the up spins, in a region in which the
contribution of the down spins in negligible, and the high temperature
shift of the maximum of the contribution of down spins.
A similar double peak is also visible in the heat capacity.
It would be interesting to investigate whether such a double peak is
also present in other spin gap systems. It is known that a double peak
is present in the specific heat of zig-zag spin ladder in a  magnetic
field.\cite{maeshima_zigzag}
To summarize,  the sum rule (\ref{eq:sumrule}) for the thermal Drude
weight,  holds for all spin systems which can be
described by an effective theory of non-interacting fermions.

\subsection{The Massive Boson
  Model}\label{sec:pure-boson}
We now consider the massive triplet boson model which was proposed as a
phenomenological model for the spin-1 Heisenberg chain. This model can
be obtained from the nonlinear sigma model\cite{haldane_gap} that
describes integer spin$-S$
chains in the limit $S\to \infty$ by softening the constraint on the $O(3)$ fields.
More precisely, this model is characterized by a Hamiltonian density \cite{affleck_field}
\begin{eqnarray}
  \label{eq:massive-bosons}
  {\cal H}(x)= \frac u 2 \sum_{\alpha=1}^3 \left[ \Pi_\alpha^2 +(\partial_x
      \phi_\alpha)^2\right] + u V({\bf \bphi}),
\end{eqnarray}
\noindent where $V(\bphi )=\frac{\Delta^2} {2u^2} {\bf \bphi^2} +
\frac{\lambda}{4}({\bf \bphi^2})^2$, and
$[\phi_\alpha(x),\Pi_\beta(x')]=i\delta(x-x')\delta_{\alpha,\beta}$.
The energy current takes the simple form \cite{li_thermal}
\begin{eqnarray}
  \label{eq:boson-thermal-current}
  j_e(x)=-u^2 \sum_{\alpha=1}^3 \Pi_\alpha \partial_x\phi_\alpha,
\end{eqnarray}
\noindent Note that this current is  independent of the potential
$V(\bphi)$ and up to a prefactor, it is just the  momentum density of
the boson field\cite{messiah_field_chapter}.  Consequently,  translation
invariance implies that the total thermal current $J_e$ is conserved.
 This allows us to use the
Eq.~(\ref{eq:drude-thermal}) to obtain the thermal Drude weight. Since
the bosons are weakly interacting, we can consider the case
$\lambda=0$,  to obtain the Drude weight
\begin{eqnarray}
  \label{eq:boson-weight}
  \tilde{\kappa}(T)=\frac 3 {8T^2} \int_{-\Lambda}^{\Lambda} dk \frac {u^4
    k^2}{\sinh^2\left(\frac{\sqrt{(uk)^2+\Delta^2}}{2T}\right)}
\end{eqnarray}
As before, the limit $T\ll \Delta$  again leads to the result
(\ref{eq:tcond-low}) for $\tilde{\kappa}(T)$ and for  $T\gg
\Delta$, we recover a linear weight
$\tilde{\kappa}(T)= \pi^2 u T$. We note that in the case of
the Takhtajan-Babujian spin-$S$ chains (which are described by $SU(2)_{2S}$ WZNW
models at low energy\cite{affleck_strongcoupl}), the weight is given
by:
\begin{equation}
\tilde{\kappa}(T)=\frac{\pi^2}{3} \frac{3S}{S+1} u T,
\end{equation}
for $S\to \infty$, this weight is the same as the one of the triplet of
bosons. This is consistent with the fact that the non-linear sigma
model describes spin$-S$ chains in the limit $S\to \infty$.

\subsection{Discussion}

In the preceding sections, we have seen that all the spin gap systems
studied in this paper exhibit the same generic behavior for the
thermal conductivity.
The reason is that for gapped 1D systems that can be bosonized, the low
energy theory is Lorenz invariant, and
excitations are described by massive particles having  relativistic-like
 dispersions $\epsilon_\alpha(p)=\sqrt{(vp)^2+m_\alpha^2}$, with
the gap $\Delta$ being the mass of the lightest particle. 
When these excitations are spin triplets, the lowest excited state
contains exactly one of these particles, and the total energy current
is 
$\epsilon(p) \frac {\partial \epsilon}{\partial p}$ which then yields 
a weight: 
\begin{eqnarray}
  \label{eq:relativistic-kappa}
  \tilde{\kappa}(T)\sim \int dp \epsilon(p)^2 \left(\frac {\partial
      \epsilon}{\partial p}\right)^2 e^{-\epsilon(p)/T}
\end{eqnarray}
Since Lorentz invariance dictates that  $\epsilon(p) \frac {\partial
  \epsilon}{\partial p}=p$, one obtains the same thermal weight as in
(\ref{eq:tcond-low}) in the low-temperature regime.   Examples of
systems possessing this
triplet branch are the alternating spin-$\frac12$ chain\cite{haldane_dimerized,barnes_dimerized}, the two-leg spin
ladder\cite{barnes_ladder,knetter_ladder} and the Heisenberg spin-1
chain\cite{takahashi_spin1}. We therefore, expect that the above
mentioned
systems  will exhibit 
a finite thermal Drude weight.  However,   this result could differ in
the case of the zig-zag ladder or the frustrated spin $\frac12$ chain.
 This stems chiefly from the fact that  though the zig-zag ladder has a 
gapful
spectrum, the low energy excitations having a relativistic dispersion, are
spinons\cite{haldane_dimerized,allen_spinons} carrying a spin
$\frac12$.
Another example with spinonic excitations is   the  
 XXZ chain in the Ising
 phase\cite{nijs_equivalence}. Since the total spin of the
system can only vary by an integer,  the spinons occur in pairs.
 Consequently, the
interaction between these spinons has a strong influence on the
thermal weight. In the case of the XXZ chain, since the spinons are
non-interacting, the current of a given excited state is conserved,
and one expects to recover  a finite Drude weight. However, in the case of
the zig-zag ladder or the frustrated spin $\frac12$ chain, the interaction
between the spinons can lead to a non-conservation of the current of
the two spinon state,  resulting  in the suppression of the thermal Drude
weight\cite{heidrich_frustrated}.   

 It would be worthwhile to compare our
predictions for the Drude weight  for various systems  with numerical
 simulations\cite{heidrich_frustrated}  or with other analytical
 techniques on the lines of  
 Ref.~\onlinecite{kluemper_xxz} in the case of  integrable models.
However, in the former case,
the  extraction of  the power law prefactor in
the activated thermal Drude weight from  numerical data
might prove very difficult.

\section{Effect of impurities}\label{sec:disorder-case}
We have seen in  Sec.~\ref{sec:pure-case} that
 the thermal conductivity in clean
systems  has a Drude peak as a result of the
translational invariance of the system. In a real system, we expect
this Drude peak to be replaced by a finite thermal conductivity, due
to the finite lifetime of eigenstates of the Hamiltonian induced by
phonon or impurity scattering.  In the present section,
we  study  the effect of impurity scattering on the gapped systems we
discussed in sections~\ref{sec:pure-ladder} and~\ref{sec:pure-XY}.
We will begin with a calculation of the conductance of the system with
a single impurity, and then we will turn to a system with a nonzero
concentration of impurities.

\subsection{Single-impurity problem}

The thermal conductivity of a system with  a single-impurity  can be
calculated  using the simple Landauer approach\cite{landauer_formula} provided,
the elementary excitations are non interacting.
The basic idea\cite{fazio_thermal_1d}
is to consider two reservoirs at temperature $T_1$ and
$T_2$ (with $T_1>T_2$) in presence of a barrier (the impurity
potential). Reservoir 1 emits a particle with momentum $k>0$, energy
$\epsilon (k)$ and velocity $\partial \epsilon (k)/\partial k$. The
probability to traverse the barrier is given by the square of the
transmission coefficient $|t(k)|^2$.  The current flowing from
reservoir 1 to reservoir 2 is:

\begin{equation}
\label{curr1}
J_{1\to 2}=\int_0^\infty  \frac{dk}{2 \pi} n_1(k,T_1) |t(k)|^2
\epsilon(k) \frac{\partial \epsilon (k)}{\partial k},
\end{equation}

\noindent and similarly the current flowing from to reservoir 2 to 1 is:

\begin{equation}
\label{curr2}
J_{2\to 1}=\int_0^\infty  \frac{dk}{2 \pi} n_2(k,T_2) |t(k)|^2
\epsilon(k) \frac{\partial \epsilon (k)}{\partial k},
\end{equation}

\noindent where  $n_{1,2}(k,T_{1,2})$ are the fermion distribution functions at
temperature $T_{1,2}$. In the limit $T_1\simeq T_2$, the net current
flowing through the barrier is:

\begin{eqnarray}
J&=&J_{1\to 2}-J_{2\to 1}=\int_0^\infty  \frac{dk}{2 \pi}
\frac{|t(k)|^2 \epsilon^2(k)}{4 \cosh (\frac{\epsilon(k)}{2 T_2})}\frac{T_1-T_2}{T^2_2} \frac{\partial \epsilon
(k)}{\partial k}\nonumber \\&=&{\cal K}(T_1)(T_1-T_2),
\end{eqnarray}

\noindent Hence a knowledge of the transmission probability $\vert
t^2\vert$
 permits us to obtain the  thermal conductance

\begin{equation}
\label{tsi}
{\cal K}(T)=\int_0^\infty \frac{dk}{2 \pi} |t(k)|^2 \frac{
\epsilon^2(k)}{4 T^2\cosh (\frac{ \epsilon(k)}{2 T})}
\frac{\partial \epsilon (k)}{\partial k},
\end{equation}
 We now  apply this general formula~(\ref{tsi})
 to two spin gap systems in
which the elementary excitations are non-interacting.
\subsubsection{Ladder with a defect}\label{sec:1imp-ladder}
We consider a two leg spin 1/2 ladder with a defect on a rung,
described by the Hamiltonian:
\begin{equation}
  \label{eq:ham-ladder-1imp}
  H=J_\parallel \sum_{n\atop p=1,2} {\bf S}_{n,p}\cdot {\bf S}_{n+1,p}
  + J_\perp \sum_{n \neq 0}  {\bf S}_{n,1}\cdot {\bf S}_{n,2} +
  J_\perp' {\bf S}_{0,1}\cdot {\bf S}_{0,2}
\end{equation}
This Hamiltonian can be fermionized following
Ref.~\onlinecite{shelton_spin_ladders}. The perturbation to the ladder
becomes:
\begin{eqnarray}
  \label{eq:perturbation}
  (J_\perp'-J_\perp)a^2  ({\bf J}_1 + {\bf n}_1)(0)\cdot ({\bf J}_2 +
  {\bf n}_2)(0),
\end{eqnarray}
\noindent where ${\bf J}_{1,2}$ and  ${\bf n}_{1,2}$ are the
uniform and staggered spin densities, respectively,  and the most
relevant contribution is $(J_\perp'-J_\perp) a^2 {\bf
  n}_1(0)\cdot {\bf n}_2(0)$. This contribution can be fermionized, so
that the resulting low energy
Hamiltonian of the ladder with a rung defect reads:
\begin{eqnarray}
  \label{eq:ladder-1imp-majorana}
  H&=&-\frac{iv}{2} \sum_{a=0}^4 \int dx (\xi_R^a \partial_x \xi_R^a -\xi_L^a
  \partial_x \xi_L^a)\nonumber \\ && + i \int dx m(x) \left(\sum_{a=0}^3
    \xi_R^a\xi_L^a -3 \xi_R^a\xi_L^a\right),
\end{eqnarray}
\noindent where $m(x)=m + g \delta(x)$, with $m=J_\perp/(2\pi)$
and $g=(J_\perp'-J_\perp)a/(2\pi)$. Clearly, each Majorana
mode is scattered independently from the barrier, so that their
contributions to the thermal conductivity is additive. As discussed in
in the preceding sections and in App.~\ref{app:weight}, we use the
correspondence between the Majorana and Dirac fermions to calculate
the thermal conductivity with the barrier. The First Quantized
Hamiltonian for the Dirac fermions reads:
\begin{eqnarray}
  \label{eq:ladder-1imp-dirac}
  H=-iv \sigma_3 \partial_x +  m(x) \sigma_2,
\end{eqnarray}
\noindent where $\sigma_i$ are Pauli matrices. Solving the  Schr{\"o}dinger
equation
with appropriate boundary conditions for the wavefunction at the barrier,
we obtain the transmission probability
\begin{equation}
\label{eq:transmission-dirac}
|t(k)|^2=\cos^2 \psi \frac{k^2}{k^2+K^2},
\end{equation}

\noindent where $K=m/v \sin \psi$ and
$\tan(\frac{\psi}{2})=\frac{g}{2v}=(J_\perp'-J_\perp)/(2\pi^2 J_\parallel)$,
Eq.(\ref{tsi}) becomes:
\begin{eqnarray}
{\cal K}(T)&=&\int_0^\infty \frac{dk}{2 \pi} \cos^2 \psi
\frac{k^2}{k^2+K^2} \frac{(vk)^2+m^2}{4 T^2\cosh (\frac{
\sqrt{(vk)^2+m^2}}{2T})} \nonumber \\ && \times\frac{(vk)}{\sqrt{(vk)^2+m^2}},
\end{eqnarray}
\noindent where we have used  $\epsilon(k)=\sqrt{(vk)^2+m^2}$.
In the limit $T\to 0$, the transmission probability is dominated by
momentum $k\ll K$ for which $|t(k)|^2\sim k^2/K^2$ i.~e. the barrier
is a strong scatterer, and
\begin{equation}\label{eq:conductance-ladder-low}
{\cal K}(T) = \frac {3m} {2\pi} e^{-m/T} \cot^2 \psi  ,
\end{equation}
\noindent where we have taken into account the  triplet of Majorana modes .
One can obtain an estimate of the temperature $T^*$ below which this result
is valid by noting that for $T\ll m$, one has $\langle (vk)^2\rangle =
mT$, so that the criterion for low temperature is $mT\ll m^2 \sin^2
\psi$, i. e. $T\ll T^*=J_\perp (J_\perp'-J_\perp)^2/J_\parallel^2$. This
temperature is clearly much smaller than the gap $m$. For higher
temperatures, $T^*\ll T \ll m$, the thermal conductance is obtained by
making the approximation $|t(k)|^2\sim \cos^2 \psi$, leading to:
\begin{eqnarray}
  {\cal K}(T)&=& \frac {3\cos^2 \psi}{4\pi}   T \int_{m/T}^\infty dx
  \left( \frac {x/2}{\cosh (x/2)}\right)^2 \nonumber \\ 
&&+\frac {\cos^2 \psi}{4\pi}
  T \int_{3 m/T}^\infty dx
  \left( \frac {x/2}{\cosh (x/2)}\right)^2
\end{eqnarray}
For $T^*\ll T \ll m$, one finds ${\cal K}(T) \sim \frac{3m^2}{4\pi T}
e^{-m/T}$, and for $T\gg m$, ${\cal K}(T)=\frac{2\pi T}{3} \cos^2
\psi$.
Contrary to the result (\ref{eq:tcond-low}) for the pure ladder, the
thermal conductance (\ref{eq:conductance-ladder-low}) for $T\ll T^*$  is purely
activated
without any 
$T$ dependent prefactor.
Therefore, the Drude weight in the thermal conductivity for the pure
system is not an accurate indication on the behavior of the thermal
conductivity in a system with impurities. The reason for that  is clear from
(\ref{eq:transmission-dirac}), namely low energy modes experience
 much stronger impurity scattering than the high-energy ones. It is
 only in the high temperature limit $T\gg m$ that the replacement
 $\delta(\omega) \to \tau$ is justified. We will see in the following
section that this result is not restricted to the spin ladder.

\subsubsection{XY chain with a defect}\label{sec:1imp-XY}

We consider again the XY-chain with alternating exchange of
Sec.~\ref{sec:pure-XY}.
We now suppose that the bond
strength $J_1$ between the sites $0$ and $1$ is  replaced by  $J'_1$.
This bond acts as a barrier and using the results of
Appendix~\ref{app:eigen-XY},  the transmission probability across this
barrier is given by
\begin{equation}
  \label{eq:transmission-XY}
  |t(k)|^2=\frac{4J_1^2 (J'_1)^2 \sin^2 \phi_k}{(J_1^2-(J'_1)^2)^2+ 4J_1^2
    (J'_1)^2 \sin^2 \phi_k}
\end{equation}

In particular, we can show that when $k\simeq\pi/2$ we have:
\begin{equation}
  |t(k)|^2=\frac{16 J_1^2 (J'_1)^2
    J_2^2 (k-\pi/2)^2}{(J_1^2-(J'_1)^2)^2(J_1-J_2)^2+ 16 J_1^2 (J'_1)^2 J_2^2 (k-\pi/2)^2},
\end{equation}
\noindent which indicates that for low temperatures $T\ll \sqrt{J_1 J_2}$,
the behavior of the
thermal conductivity in the XY chain with a bond defect is identical
to the behavior of the thermal conductivity in the ladder
discussed in Sec.~\ref{sec:1imp-ladder}.
For high temperatures, $T\gg
\sqrt{J_1 J_2}$, we can neglect the variation of the transmission
coefficient with the energy, and assume that all states have the same
probability of occupation. Then, the thermal conductance reads:
\begin{equation}
  {\cal K}(T)=T \int_{(J_1-J_2)/T}^{(J_1+J_2)/T} d\epsilon \epsilon^2
  \langle |t(\epsilon)|^2\rangle \sim 1/T^2
\end{equation}

\subsection{Many impurities case}
In this section, we consider the effect of a finite concentration of
impurities on the thermal conductivity of the ladder. As before,  the disorder we consider is a
random rung coupling. The Hamiltonian of the disordered ladder reads:
\begin{eqnarray}
 H=J_\parallel \sum_{i\atop {p=1,2}} {\bf S}_{i,p} \cdot {\bf
  S}_{i+1,p} +  \sum_i J^i_\perp {\bf S}_{i,1}\cdot {\bf S}_{i,2},
\end{eqnarray}
\noindent where $J_\perp^i= J_\perp+\eta_i$. We have $\vert\eta_i\vert<J_\perp$,
  so that all rung interactions remain antiferromagnetic.
  This Hamiltonian can be analyzed by mapping onto a random
mass Majorana fermions
model\cite{gogolin_disordered_ladder,steiner_random_mass}.

\begin{eqnarray}
H&=&-\sum_{a=1}^4 \int dx \{ \frac{iv}{2}\lbrack \xi^a_R(x) \partial_x
\xi^a_R(x)-\xi^a_L(x) \partial_x \xi^a_L(x) \nonumber \\ &&+im^a(x)
\xi^a_R(x)\xi^a_L(x) \rbrack \},
\end{eqnarray}

\noindent with  $m^{1,2,3}(x)=m(x)$ for
the triplet magnetic excitation, $m^0=-3|m|(x)$ for the singlet
excitation and $m(x)=m+\eta(x)$ where
$\overline{\eta(x)\eta(x')}=D\delta(x-x')$. We note that disorder
does not mix the different flavors of Majorana fermions. Consequently,
the contribution of the Majorana modes to the  thermal conductivity
remains additive. As before, to calculate the
disorder induced self-energy, it is useful to  recast the above
problem in terms of Dirac Fermions.
\begin{eqnarray}
\label{eq:random_dirac}
H&=&-i v \int dx (\psi^\dagger_R \partial_x \psi_R - \psi^\dagger_L
\partial_x\psi_L) \nonumber \\ &&+ m(x) \int dx (\psi^\dagger_R\psi_L +
\psi^\dagger_L\psi_R),
\end{eqnarray}
We note that the Hamiltonian (\ref{eq:random_dirac}) can also be
derived from a dimerized XY chain with bond
defects.\cite{mckenzie_random_xy}
We define the $2\times2$ matrix disordered averaged Green's function $\hat{G}$ by its components,
\begin{equation}
G_{\alpha \beta}(x,\tau)= - \langle T_\tau \psi_\alpha(x,\tau)
\psi^\dagger_\beta(0,0) \rangle,
\end{equation}
\noindent where $\alpha,\beta \in \{R,L\}$.
The Hamiltonian (\ref{eq:random_dirac}) can be rewritten in matrix
form as:
\begin{equation}
H=\int dx \Psi^\dagger(x)[-i v \tau_3 \partial_x + m(x) \tau_1]\Psi(x) ,
\end{equation}
\noindent where $\tau_{1,3}$ are Pauli matrices.
The impurity self-energy matrix $\hat{\Sigma}$ can be  calculated within
the Self Consistent Born Approximation (SCBA)\cite{mori_scba}  and
satisfies the  Dyson equation for
the disorder averaged Green's function:
\begin{equation}\label{eq:self}
(i \omega_n - vk \tau_3 - m\tau_1 - {\Sigma}){G} = 1,
\end{equation}
 In
this approximation, the self-energy is independent of momentum  and is
determined self-consistently by
\begin{equation}
{\Sigma(i\omega_n)} = D \int \frac{dk}{2\pi} \tau_1 [i\omega_n - v
k \tau_3 -  m\tau_1  -{\Sigma(i\omega_n)}]^{-1}\tau_1
\end{equation}
Clearly,  the self energy possesses the  following structure,
${\Sigma}(i\omega_n) = i \sigma(i\omega_n) + V(i\omega_n)\tau_1$
leading  to the following self-consistent equations for $\sigma$ and $V$\cite{mori_scba}:
\begin{eqnarray}\label{eq:scba}
\sigma(i\omega_n)= \frac{D}{2v}
\frac{\sigma(i\omega_n)-\omega_n}{\sqrt{(\sigma(i\omega_n)-\omega_n)^2
+ (m + V(i\omega_n))^2}} \nonumber \\
V(i\omega_n)=\frac{D}{2v}
\frac{V(i\omega_n) -m}{\sqrt{(\sigma(i\omega_n)-\omega_n)^2 +
(m + V(i\omega_n))^2}}
\end{eqnarray}
Introducing the dimensionless variables:
$s=i\sigma/m$, $t = V/m$, $x= i\omega_n/m$ and $\lambda= D/(2vm)$
the above self-consistent equations  simplify to:
\begin{eqnarray}
&& t=\frac{s}{x-2s}, \\ && s^4 -s^3 x - s^2 (\frac{1-x^2} 4 -
\lambda^2) - \lambda^2 sx + \frac{\lambda^2} 4 x^2 =0.
\end{eqnarray}
\begin{figure}[htbp]
  \begin{center}
\centerline{\includegraphics[angle=-0,width=\normwidth]{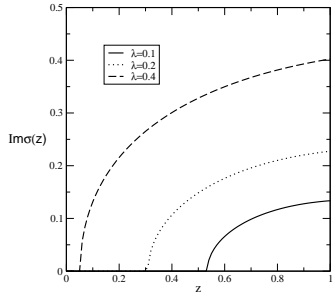}}

    \caption{The imaginary part of the self energy in units of $m$
calculated from
      Eq.~(\ref{eq:scba}) for various values of the scaled impurity
strength $\lambda$. Note that the effective gap decreases with increasing
       $\lambda$.}
    \label{fig:self-energy}
  \end{center}
\end{figure}

\begin{figure}[htbp]
  \begin{center}
\centerline{\includegraphics[angle=-0,width=\normwidth]{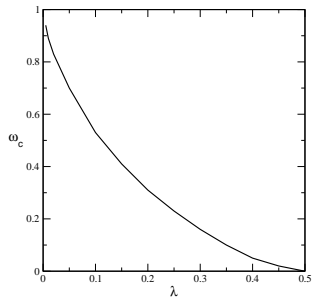}}

\caption {The effective gap in units of $m$ as a function of the
scaled impurity strength $\lambda$.}
\label{fig:graph3}
  \end{center}
\end{figure}

This quartic equation can be solved numerically. Some sample curves
are shown in Fig.\ref{fig:self-energy}. We find that the disorder renormalizes
the
gap in the spectrum and a sufficiently strong disorder ($\lambda=0.5$)
 closes the
gap indicating a disorder induced phase transition within the SCBA.
A plot of the renormalized gap
$\omega_c$ as a function of disorder strength is shown on Fig.~\ref{fig:graph3}.
In the ensuing calculation, we only consider disorder strengths for which
the renormalized gap is non-zero.

\begin{figure}[htbp]
  \begin{center}
\centerline{\includegraphics[angle=-0,width=\normwidth]{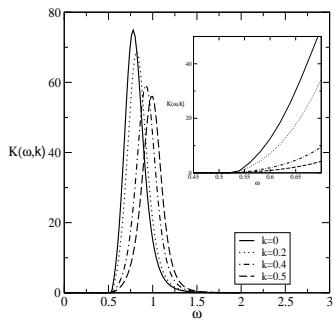}}
     \caption{The function $K(\omega,k)$ in units of $m^{-2}$ for
$\lambda=0.1$ and various values of $k$ (in units of $m/v$.)}
    \label{fig:graph2}
  \end{center}
\end{figure}

Using the results of the previous sections, the thermal conductivity
for the disordered ladder can be rewritten as \cite{mahan_book}:
\begin{equation}
\label{eq:mahan_formula}
\kappa(T)=\frac 1 {T} \int \frac{d\omega }{2\pi}
\left(-\frac{\partial n_F}{\partial \omega }\right) [
P(\omega-i0,\omega+i0) -\mathrm{Re} P(\omega+i0,\omega+i0)],
\end{equation}
\noindent where:
\begin{equation}\label{eq:pww}
P(\omega,\omega')=\int \frac{dk}{2\pi} (v^2k)^2 \mathrm{Tr}\left[
G(k,\omega) G(k,\omega')\right],
\end{equation}
\noindent Vertex corrections to (\ref{eq:mahan_formula}) are
negligible
in the weak disorder limit.
Contrary to the suggestion in Ref.~\onlinecite{mahan_book} that
$\mathrm{Re} P(\epsilon+i0,\epsilon+i0)$ in
Eq. (\ref{eq:mahan_formula}) can be neglected, we find that this term
is indeed crucial to take into account the presence of a gap in the
energy spectrum. To proceed with the calculation of $\kappa$, we first note
that for weak disorder i.e., $D\ll v m$,  since the off-diagonal
self-energy $V$ always occurs in the combination $m+V$
(\ref{eq:self}), it is reasonable to
neglect $V$ in the Green's function $G$ which can then be approximated
as
\begin{equation}
G(k,\omega )=\frac{\omega -\sigma(\omega) +vk \sigma_3 +m \sigma_1}{[\omega -\sigma(\omega)]^2
-\epsilon(k)^2},
\end{equation}
\noindent
where $\epsilon(k)=\sqrt{v^2k^2 + m^2}$. This yields
\begin{eqnarray}
G(k,\omega+i0)&=&\left(\frac 1 2 + \frac{vk \sigma_3 +
m\sigma_1}{2\epsilon(k)}\right) \frac 1
{\omega-\sigma(\omega)-\epsilon(k)}\nonumber \\ && + \left(\frac 1 2 - \frac{vk
\sigma_3 + m\sigma_1}{2\epsilon(k)}\right) \frac 1
{\omega-\sigma(\omega)+\epsilon(k)},\nonumber \\
G(k,\omega-i0)&=&\left(\frac 1 2 + \frac{vk \sigma_3 +
m\sigma_1}{2\epsilon(k)}\right) \frac 1
{\omega-\sigma^*(\omega)-\epsilon(k)}\nonumber \\ && + \left(\frac 1 2 - \frac{vk
\sigma_3 + m\sigma_1}{2\epsilon(k)}\right) \frac 1
{\omega-\sigma^*(\omega)+\epsilon(k)}.\nonumber
\end{eqnarray}
Substituting the above in (\ref{eq:pww}),
we obtain
\begin{equation}
P(\omega+i0,\omega-i0)-\mathrm{Re}P(\omega+i0,\omega+i0) = \int
\frac{dk}{2\pi} (v^2k)^2 K(\omega,k)
\end{equation}
\noindent where
\begin{eqnarray}
  \label{eq:definition_K}
 K(\omega,k)&=&\frac{(\mathrm{Im} \sigma(\omega))^2}{\left[
(\omega - \epsilon(k) -\mathrm{Re}\sigma(\omega))^2+
(\mathrm{Im}\sigma(\omega))^2\right]^2} \nonumber \\ &+& \frac{(\mathrm{Im}
\sigma(\omega))^2}{\left[ (\omega + \epsilon(k)
-\mathrm{Re}\sigma(\omega))^2+
(\mathrm{Im}\sigma(\omega))^2\right]^2},
\end{eqnarray}
\noindent
This expression can now be used in (\ref{eq:mahan_formula}) to obtain
the
thermal conductivity.
At low temperatures,  the derivative of the
Fermi function in (\ref{eq:mahan_formula}) decays exponentially as
$\exp-(\omega/T)$ indicating that frequencies much larger than
$T$ can be neglected in the integral for the thermal conductivity
$\kappa(T)$.
Consequently,  the low temperature
behavior of $\kappa (T)$  is completely dictated by the
 the low frequency behavior of
$K(\omega ,k)$.  We now analyze the behavior of $K$. Firstly,   since the
diagonal self energy $\mathrm{Im}\sigma(\omega+i0)=0$, for  $\omega< \omega_c$ i.e., for
frequencies smaller than the disorder renormalized gap,
$K(\omega ,k)$ is identically zero for all $\omega< \omega_c$ .  A typical plot of $K$
as a function of $\omega $ for two different values of $k$ is shown
on figure~\ref{fig:graph2}.

\noindent
Clearly, the dominant contribution to $\kappa$ for temperatures $T<\omega_c$
comes from the behavior of
$K(\omega ,k)$  in the vicinity of $\omega_c$. This behavior has been
analyzed numerically, and we find that for all $k$, $K$ can be
developed as a series in $\omega-\omega_c$:
\begin{eqnarray}
  \label{eq:fit-K}
  K(\omega,k)&=&\alpha(k)(\omega-\omega_c)\nonumber \\ 
&& +\beta(k)(\omega-\omega_c)^2+\ldots
\end{eqnarray}
We find that $\alpha,\beta$ are fast decreasing functions of $|k|$,
such that $\int dk k^2 \alpha(k) <\infty$ and $\int dk k^2 \beta(k)
<\infty$. Substituting (\ref{eq:fit-K}) in (\ref{eq:mahan_formula}),   we obtain
\begin{eqnarray}
  \kappa(T)&=&\frac 1 {4T^{2}} \int dk \int d\omega \; k^2
  e^{-\omega/T}\Theta(\omega-\omega_c)
  \left[\alpha(k)(\omega-\omega_c)\right. \nonumber \\ &&\left. +\beta(k)
    (\omega-\omega_c)^2+\ldots \right]
\end{eqnarray}
leading to,
\begin{equation} \label{eq:kappa}
 \kappa(T) \sim \tilde{\alpha}
e^{-\omega_c/T}+ \tilde{\beta} T e^{-\omega_c/T} + o(T
e^{-\omega_c/T})
\end{equation}
\noindent $\tilde{\alpha}, \tilde{\beta}$ are
temperature independent constants. We find that the results are
similar to those for a  single impurity
(\ref{eq:conductance-ladder-low})
with the difference that
a  finite concentration of
impurities renormalizes  the gap.
At very high temperatures one recovers the usual $T^{-2}$ decrease of
the
thermal conductivity. It would be interesting to obtain the cross-over
behavior from the the low to high temperature regimes. However, the
rather
complicated form of the self energies  makes analytical calculations
very difficult for these intermediate temperatures and this is left
for future work.
 For small magnetic fields $h\ll \omega_c$ the
same result holds with the substitution $\omega_c \to \omega_c
- h$.

\section{Discussion}

We now highlight the connection between our approach and that of the
Boltzmann equation. A Boltzmann like equation for the thermal
conductivity
can be recovered from the SCBA   in  the limit of small self
energies.
For  $\sigma(\omega )\ll m$,
  the function $K$ defined by
(\ref{eq:definition_K}), takes the simpler form
\begin{equation}
  \label{eq:boltzmann}
  K(\omega ,k)=\pi [\delta(\omega-\epsilon(k))
  +\delta(\omega-\epsilon(k))] (\rm{Im} \sigma(\omega))^{-1}
\end{equation}
Inserting the above result in  (\ref{eq:mahan_formula}), we obtain:
\begin{equation}
  \label{eq:kappa-boltzmann}
  \kappa(T)=\frac 1 {T} \int \frac{dk}{2\pi} \left(\epsilon(k)
  \frac{\partial\epsilon(k)}{\partial k}\right)^2\left(-\frac{\partial
      n_F}{\partial \epsilon}\right)(\epsilon(k))  (\rm{Im} \sigma(\epsilon(k)))^{-1}
\end{equation}
We see that in the limit of very small self energies, we recover the
 Boltzmann equation
result for the thermal conductivity\cite{ziman_solid_book}. Comparing
(\ref{eq:kappa-boltzmann}) with (\ref{eq:drude-thermal}) and
(\ref{eq:ladder-kappa}),
we see that if $Im \sigma(\epsilon(k))= \tau^{-1}$ where
$\tau$ is a constant
independent of $\epsilon$, the thermal conductivity can be written as a
product of the Drude weight in the absence of impurities and the mean
relaxation
time $\tau$ as was proposed in Ref.\onlinecite{alvarez_ladder}.
The underlying assumption there, was that
all the  eigenstates of the system  have the same lifetime $\tau$
independent of the energy of the eigenstate.
As shown above,  the explicit energy dependence of the self
energy  found in Sec.~\ref{sec:disorder-case}, i.e., the fact that the low
energy spin excitations are much more scattered by impurities than
high energy excitations,
shows that any assumption of energy independent lifetime
is invalid even for the simplest models. As a result, the thermal
Drude weight can  atmost  yield a heuristic  behavior
of the thermal conductance in a real system in which spin excitations
are interacting with impurities and/or phonons due to the different
extrinsic lifetimes of current carrying states.

 We present a brief comparison of our results with experiments. Since disorder
is ubiquitous in real systems,  it is
reasonable to compare our results for the two  leg ladder with
impurities  with experimental measurements.
One of the systems studied extensively is the spin gap compound
$\mathrm{Sr_{14-x}(La,Ca)_xCu_{24}O_{41}}$
\cite{sologubenko_thermal_ladder,hess_thermal_ladder,kudo_thermal_ladder}.
In the insulating phase, these systems can be well
described by an array of  two-leg spin ladders\cite{uchara_SrCaCuO}.
 In this system,
the phonon subtracted thermal conductivity was shown to have an
exclusive
spin contribution. At low temperatures, a fit for  spin thermal conductivity
yielded a  $\kappa(T) \sim e^{-\Delta/T}$. This low temperature fit is in good
accord with our
prediction of
$\kappa(T) \sim e^{-\omega_c/T}$. However, a full quantitative comparison requires an
understanding of the effect of the disorder in the material on the
spins, the effect of spin-phonon interactions and other exchange
interactions in the ladder.

\section{Conclusion}

In the present paper, we have calculated the thermal conductivity of
 gapless spin chains and  spin gap systems  including 
 the two-leg spin-$\frac12$
ladder and the dimerized XY spin-$\frac12$ chain. In the absence of
disorder,
the thermal Drude weight of gapless spin chains vanishes linearly with
temperature. On the other hand, for the ladder and the XY chain, which
can both be
represented as free fermions, we find that the thermal 
 Drude weight  $\tilde{\kappa} \sim T^{-1/2} e^{-m/T}$, where $m$ is the gap
to the lowest triplet excitation.  For intermediate temperatures,
${\tilde \kappa}
\propto T$ and decays as $T^{-2}$ at very high temperatures. We argue that
this
behaviour is generic to all quasi-one dimensional spin gapped systems
having low energy triplet excitations with a relativistic-like
dispersion. We have also considered the effect of
a 
 magnetic field which
results in a substantial enhancement of the low temperature thermal
conductivity. Furthermore, in the case of dimerized XY chains, a double
peak can be obtained in the thermal conductivity for large enough
fields. We have also studied the effect of  impurities on spin gap
systems like the ladder and the XY chain. 
Impurities destroy the Drude peak
resulting in  a finite thermal conductivity at zero frequency. This
thermal conductivity has a generic form $e^{-\tilde{m}/T}$ at low temperatures
 where, $\tilde{m}$ is the effective gap of the
system.  It would be interesting to include the effects of magnetic
impurities and also scattering from phonons. These and other questions
are left for future work.

\section{Acknowledgments}
We thank the authors of Ref.\onlinecite{heidrich_frustrated} for their
comments on a previous version of this manuscript. 
R.C. acknowledges {\'E}cole Normale Sup{\'e}rieure for kind hospitality during
the completion of the present work.

\appendix
\section{Calculation of the thermal conductivity for Majorana
  fermions}\label{app:weight}
\subsection{Thermal conductivity of Dirac fermions}
We consider massive Dirac fermions with the following Hamiltonian density:
\begin{eqnarray}\label{eq:hamiltonian-dirac}
{\cal H}(x)&=&-i \frac v 2 (\psi^\dagger_R \partial_x \psi_R -
(\partial_x \psi_R^\dagger) \psi_R  - \psi^\dagger_L
\partial_x\psi_L +(\partial_x\psi_L^\dagger)\psi_L ) \nonumber \\ &&
 + m (e^{i\varphi} \psi^\dagger_R\psi_L +e^{-i\varphi} \psi^\dagger_L\psi_R),
\end{eqnarray}
In this case, the energy current reads:
\begin{equation}
  \label{eq:current-dirac}
  j_e(x)=-i \frac{v^2} 2 (\psi^\dagger_R \partial_x \psi_R -
  (\partial_x \psi^\dagger_R) \psi_R + \psi^\dagger_L \partial_x
  \psi_L - (\partial_x \psi^\dagger_L) \psi_L)(x)
\end{equation}
We note that using the  transformation $\psi_L \to \varphi \psi_L$, we
can reduce the Hamiltonian (\ref{eq:hamiltonian-dirac}) to the case
$\phi=0$, while leaving the current (\ref{eq:current-dirac})
invariant. Therefore, for the purpose of calculating the thermal
transport we can without loss of generality restrict to the case
$\phi=0$ in (\ref{eq:hamiltonian-dirac}).

Using the Fourier decomposition
\begin{eqnarray}
  \label{eq:fourier-dirac}
  \psi_{\nu}(x)=\frac 1 {\sqrt{L}} \sum_k c_{k,\nu} e^{ik x}
\end{eqnarray}
 the  Dirac Hamiltonian $H=\int dx {\cal H}(x)$  can be diagonalized to  obtain
\begin{eqnarray}
  \label{eq:diagonal-dirac}
  H=\sum_k \epsilon(k) (c^\dagger_{k,+} c_{k,+} -c^\dagger_{k,-}
  c_{k,-})
\end{eqnarray}
where the fermionic operators $c_{k,\pm}$    are linear combinations of the
$c_{k,R/L}$
such that $c^\dagger_{k,+}
c_{k,+}+c^\dagger_{k,-} c_{k,-}=c^\dagger_{k,R} c_{k,R}
+c^\dagger_{k,L} c_{k,L}$ and $\epsilon(k)=\sqrt{(vk)^2+m^2}$.
This allows us to rewrite the total energy current $J_e= \int dx j_e(x) =\sum_k v^2 k (c^\dagger_{k,R} c_{k,R} +
c^\dagger_{k,L} c_{k,L} -1)$  as:
\begin{eqnarray}
  J_e=\sum_k v^2 k (c^\dagger_{k,+} c_{k,+} + c^\dagger_{k,-} c_{k,-}
-1).
\end{eqnarray}
\noindent
Using (\ref{eq:diagonal-dirac}), one easily obtains
\begin{eqnarray}
  \langle J_e^2 \rangle = 2 \sum_k (v^2 k)^2 \langle n_+(k) \rangle
  (1- \langle n_+(k) \rangle)
\end{eqnarray}
\noindent where the Fermi distribution function
$ \langle n_+(k) \rangle = (e^{\beta\epsilon(k)}+1)^{-1}$. From (\ref{eq:drude-thermal}), the  thermal conductivity
$\kappa(\omega,T)=\tilde{\kappa}\delta(\omega)$ with a Drude weight
\begin{eqnarray}
  \label{eq:conductivity-dirac}
\tilde{\kappa}(T)= \frac 1 {4T^2} \int_{-\infty}^{\infty} dk \frac
{v^4 k^2}{\cosh^2 \left( \frac{\epsilon(k)}{2T}\right) }
\end{eqnarray}
The above result  has a simple interpretation in terms of  kinetic
theory. Consider the expression for the specific heat:
\begin{equation}
C_v(T)=\frac 1 {T^2} \int \frac{dk}{2\pi} \frac {\epsilon(k)^2}
{\cosh^2 \left( \frac{\epsilon(k)}{2T}\right) } \equiv \int \frac{dk}{2\pi}
c_v(k,T),
\end{equation}
i.e., a mode of momentum $k$ contributes $c_v(k)$
to the specific heat. Such a mode has a velocity $v(k)=v^2
k/\epsilon(k)$. This now permits us to rewrite (\ref{eq:kappa-t}) as:
\begin{equation}\label{eq:kinetic-interpretation}
\tilde{\kappa} = \int \frac{dk}{2\pi} c_v(k,T) v^2(k),
\end{equation}
i.e., the contribution of each mode $k$ to the  the Drude weight is
just the product of its specific heat and square of the  velocity.
This is similar to  the kinetic theory result that the thermal
Drude weight is given by the product of
the
specific heat and the square of the velocity of the free modes.

\subsection{ Majorana fermions}
It is well known that the Dirac Hamiltonian (\ref{eq:hamiltonian-dirac}) can be re-expressed
in terms of two Majorana fermions fields defined by
$\psi_\nu=(\zeta_\nu^1+i\zeta_\nu^2)/\sqrt{2}$. The Hamiltonian  can be written as a
sum of two Majorana Hamiltonians
\begin{eqnarray}
H_{\text{Dirac}}=H_M[\zeta^1]+H_M[\zeta^2]
\end{eqnarray}
Similarly, the energy current (\ref{eq:current-dirac}) can be written
as the sum of two energy currents, each  associated with one
Majorana field: $j^D_e(x)=j^1_e(x)+j^2_e(x)$. The thermal conductivity
of the Dirac Hamiltonian can then be written as the sum of the
conductivities associated with the two Majorana field i.e.,
$\kappa_{\text{Dirac}}(\omega,T)=\kappa^1(\omega,T)+\kappa^2(\omega,T)$.
The expression of the currents and the Hamiltonian being identical for
the two species of Majorana fermions, it is clear that
$\kappa^1(\omega,T)=\kappa^2(\omega,T)$. Thus, one obtains the generic result that
\begin{eqnarray}
  \label{eq:majorana-conduction}
  \kappa_{\text{Majorana}}(\omega,T)=\kappa_{\text{Dirac}}(\omega,T)/2.
\end{eqnarray}
This result shows that it suffices to calculate the thermal
conductivity of the corresponding Dirac Hamiltonian to obtain the
Majorana thermal conductivity. This correspondence holds  provided there are no
interactions
between the various species of Majorana fermions.

\section{Thermal current in the presence of an applied magnetic field}
\label{app:mag-clean}
In the presence of an applied magnetic field, the Hamiltonian density
is  ${\cal H}(x)={\cal
  H}(x)^{{\bf h}=0}-{\bf h}\cdot {\bf m}(x)$, where ${\bf m}(x)$ is
the magnetization density  and ${\bf
  h}$ is the external  magnetic field. Using the continuity equation for the
Hamiltonian density and the  equation of conservation of the
moment $\partial_t {\bf m}+\partial_x {\bf j}_s=0$,
 the thermal current is now given by
\begin{eqnarray}
  \label{eq:mag-current}
  j_{\text{th.}}(x)=j_e(x)-{\bf h}\cdot {\bf j}_s(x),
\end{eqnarray}
\noindent where $j_e$ is the energy current for ${\bf h}=0$ and
 ${\bf j}_s(x)$ is the magnetization current.

For the specific case of the ladder with a magnetic field along
the $z$ direction, the Pauli coupling is
\begin{eqnarray}
  \label{eq:ladder-field}
  H_{\text{mag.}}=-ih \int dx (\xi_R^1 \xi_R^2+\xi_L^1 \xi_L^2)
\end{eqnarray}
Note  that the contribution to the thermal conductivity arising
from the $\xi^0_{R,L},\xi^3_{R,L}$ is not changed by the application
of the magnetic field. To obtain the thermal conductivity coming from
the modes $\xi^{1,2}$ it is convenient to turn to the Dirac
Fermions\cite{shelton_spin_ladders,orignac_2spinchains}
$\psi_{\nu,s}=(\xi_\nu^1+i\xi_\nu^2)/\sqrt{2}$.
Then, one can rewrite $H_{\text{mag.}}$ as:
\begin{eqnarray}
  \label{eq:ladder-field-dirac}
  H_{\text{mag.}}=-h \int dx
  (\psi^\dagger_{R,s}\psi_{R,s}+\psi^\dagger_{L,s}\psi_{L,s})
\end{eqnarray}
The expression of the total thermal current then reads:
\begin{eqnarray}
  \label{eq:thermal-current-field}
  J_e&=& \sum_k \left[ (\epsilon(k)-h) \frac{\partial \epsilon}{\partial
      k} (c^\dagger_{k,+}c_{k,+}-\langle
    c^\dagger_{k,+}c_{k,+}\rangle)\right.\nonumber \\ && \left. -(\epsilon(k)+h)\frac{\partial \epsilon}{\partial
      k} (c^\dagger_{k,-}c_{k,-}-\langle
    c^\dagger_{k,-}c_{k,-}\rangle)\right]
\end{eqnarray}
The contribution of the $\xi^{1,2}$ modes to the Drude weight in the
thermal conductivity is then calculated to be
\begin{eqnarray}
  \label{eq:drude-field}
  \tilde{\kappa}^1(T,h)+\tilde{\kappa}^2(T,h)&=&\frac 1 {8T^2}\int dk
  \left[
    \frac{(\epsilon(k)-h)^2}{\cosh\left(\frac{\epsilon(k)-h}{T}\right)}\right. \nonumber\\ && \left.  + \frac{(\epsilon(k)+h)^2}{\cosh\left(\frac{\epsilon(k)+h}{T}\right)}\right]
\end{eqnarray}

\section{Eigenvalues and Eigenstates of the fermionized XY
  chain}\label{app:eigen-XY}
\subsection{Translational Invariant Case}
The eigenstates of the Hamiltonian (\ref{eq:XY-fermionized})
 are obtained by solving the equations:
\begin{eqnarray}
  J_1 A_{2n}+J_2A_{2n+2}=E A_{2n+1} \label{eq:eig-odd} \\
J_2 A_{2n-1}+J_1 A_{2n+1}=E A_{2n}\label{eq:eig-even}
\end{eqnarray}

One finds positive energy solutions:
\begin{equation}
  \label{eq:positive-eig}
  \left(\begin{array}{c} A_{2n} \\ A_{2n+1}\end{array} \right)=
  e^{2ikn} \left( \begin{array}{c} e^{-i\phi_k/2} \\
      e^{i\phi_k/2}\end{array}\right)
\end{equation}
with $E(k)=\sqrt{(J_1-J_2)^2+4 J_1 J_2 \cos^2 k}$ and
$J_1+J_2e^{2ik}=E(k) e^{i\phi_k}$, and negative energy solutions:
\begin{equation}
  \label{eq:negative-eig}
  \left(\begin{array}{c} A_{2n} \\ A_{2n+1}\end{array} \right)=
  e^{2ikn} \left( \begin{array}{c} - e^{-i\phi_k/2} \\
      e^{i\phi_k/2}\end{array}\right)
\end{equation}
with $E(k)=-\sqrt{(J_1-J_2)^2+4 J_1 J_2 \cos^2 k}$ and
$J_1+J_2e^{2ik}=|E(k)| e^{i\phi_k}$,

\subsection{Single impurity case}
Clearly, the solutions with
momentum $k$ and $-k$ are degenerate in energy, thus we search the
solution as a linear combination of these solutions.
The system of equations to solve reads:
\begin{eqnarray}
  \label{eq:defect}
  J_1 A_{2n} + J_2 A_{2n+2}=E A_{2n+1} (n\neq 0) \label{eq:odd-gen}\\
  J_2 A_{2n-1}+J_1 A_{2n+1}=E A_{2n} (n\neq 0) \label{eq:even-gen}\\
  J'_1 A_0+J_2 A_2=E A_1 (n=0)\label{eq:odd-zero}\\
  J_2 A_{-1}+J'_1A_1=E A_0 (n=0) \label{eq:even-zero}
\end{eqnarray}

We search for solutions of the form:
\begin{equation}
  \label{eq:left-sol}
  \left(\begin{array}{c} A_{2n} \\ A_{2n+1}\end{array} \right)=
  e^{2ikn} \left( \begin{array}{c} e^{-i\phi_k/2} \\
      e^{i\phi_k/2}\end{array}\right) + r(k) e^{-2ikn} \left( \begin{array}{c} e^{i\phi_k/2} \\
      e^{-i\phi_k/2}\end{array}\right)
\end{equation}
for $n\leq -1$ and:
\begin{equation}
  \label{eq:right-sol}
 \left(\begin{array}{c} A_{2n} \\ A_{2n+1}\end{array} \right)= t(k) e^{2ikn} \left( \begin{array}{c} e^{-i\phi_k/2} \\
      e^{i\phi_k/2}\end{array}\right)
\end{equation}
for $n\geq 1$.
Applying equations (\ref{eq:odd-gen}) for $n=-1$ and
(\ref{eq:even-gen}) for $n=1$ we obtain respectively:
\begin{eqnarray}
  \label{eq:particular}
 && A_0=e^{-i\phi_k/2}+r(k) e^{i\phi_k/2} \nonumber \\
 && A_1=t(k) e^{i\phi_k/2}
\end{eqnarray}
The equations that determine $t,r$ are obtained from
(\ref{eq:odd-zero}) and (\ref{eq:even-zero}). They read:
\begin{eqnarray}
  \label{eq:rtsystem}
  && J_2(e^{-2ik} e^{i\phi_k/2} + r(k) e^{2ik} e^{-i \phi_k/2} + J'_1
  t(k) e^{i\phi_k/2} =E(e^{-i\phi_k/2}+r(k) e^{i\phi_k/2}) \\
  && J_1'(e^{-i\phi_k/2} +r(k) e^{i\phi_k/2}) +J_2 t(k) e^{2ik}
  e^{-i\phi_k/2} =E t(k) e^{i\phi_k/2}
\end{eqnarray}
Using the relation $J_1+J_2e^{i2k}=E(k)e^{i\phi_k}$ these equations
are simplified as follows:
\begin{eqnarray}
  -J_1 e^{-i\phi_k/2} r(k)+J'_1 e^{i\phi_k/2}t(k)=J_1 e^{i\phi_k/2} \\
J'_1 e^{i\phi_k/2} r(k) -J_1 e^{-i\phi_k/2} t(k) =-J'_1 e^{-i\phi_k/2}
\end{eqnarray}

We obtain the transmission amplitude $t(k)$ and the reflection
amplitude $r(k)$ as:
\begin{eqnarray}
  \label{eq:rt-coeff}
  r(k)=\frac{(J'_1)^2-J_1^2}{J_1^2 e^{-i\phi_k}-(J'_1)^2e^{i\phi_k}} \\
  t(k)=\frac{-2i J_1 J'_1 \sin \phi_k}{J_1^2 e^{-i\phi_k}-(J'_1)^2e^{i\phi_k}}
\end{eqnarray}


\end{document}